\documentclass[reprint,aps,prl,floatfix,superscriptaddress,showpacs]{revtex4-1}
\usepackage{mathtools}
\usepackage{amssymb}
\usepackage{graphicx}
\usepackage{hyperref}

\hypersetup{
  colorlinks   = true, 
  urlcolor     = blue, 
  linkcolor    = blue, 
  citecolor   = blue 
}

\begin{document}


\title{Quantum Lissajous Scars}

\author{J.~Keski-Rahkonen}
\affiliation{Computational Physics Laboratory, Tampere University, Tampere 33720, Finland}

\author{A.~Ruhanen}
\affiliation{Computational Physics Laboratory, Tampere University, Tampere 33720, Finland}

\author{E.~J.~Heller}
\affiliation{Department of Physics, Harvard University, Cambridge, Massachusetts 02138, USA}

\author{E.~R\"{a}s\"{a}nen}
\affiliation{Computational Physics Laboratory, Tampere University, Tampere 33720, Finland}
\affiliation{Department of Physics, Harvard University, Cambridge, Massachusetts 02138, USA}

\date{\today}


\begin{abstract}
A quantum scar -- an enhancement of a quantum probability density in the vicinity of a classical periodic orbit -- is a fundamental phenomenon connecting quantum and classical mechanics. Here we demonstrate that some of the eigenstates of the perturbed two-dimensional anisotropic (elliptic) harmonic oscillator are strongly scarred by the Lissajous orbits of the unperturbed classical counterpart. In particular, we show that the occurrence and geometry of these quantum Lissajous scars are connected to the anisotropy of the harmonic confinement, but unlike the classical Lissajous orbits the scars survive under a small perturbation of the potential. This Lissajous scarring is caused by the combined effect of the quantum (near) degeneracies in the unperturbed system and the localized character of the perturbation. Furthermore, we discuss experimental schemes to observe this perturbation-induced scarring.
\end{abstract}

\maketitle


The harmonic oscillator (HO) is a linchpin in various fields of physics~\cite{HO_review}. The periodic orbits (POs) of the two-dimensional (2D) \emph{anisotropic} (elliptic) HO were first investigated by Bowditch~\cite{Bowditch} and later in more detail by Lissajous~\cite{Lissajous}. These Lissajous orbits are sensitive on the frequency ratio of the confinement. In contrast, the corresponding quantum eigenfunctions possess the same rectangular symmetry as solved in terms of the Hermite-Gaussian (HG) modes~\cite{HG_modes}, regardless of the value of the frequency ratio.     

The HG modes can be experimentally studied from laser transverse modes due to the analogy of the Schr\"{o}dinger equation with the wave equation~\cite{HG_modes_probing}. On the other hand, the HO has turned out to be a suitable prototype model for semiconductor quantum dots (QDs)~\cite{QD_apllications}. However, actual QD devices are influenced by impurities and imperfections (see, e.g., Refs.~\cite{QD_and_disorder_(I), QD_and_disorder_(II),QD_and_disorder_(III),QD_and_disorder_(IV)}). If high-energy eigenstates of a generic, perturbed QD were indeed featureless and random, controlled applications in this regime would be tedious to realize. Besides additional deflects, anisotropic QDs have attracted general interest in connection with the chaotic behavior as well as the properties in an external magnetic field~\cite{QD_and_chaos,QD_and_B_(I), QD_and_B_(II),QD_and_B_(III),QD_and_B_(IV),QD_and_B_(V),QD_and_B_(VI)}.

Nonetheless, in consequence of quantum interference, the probability density of a quantum state can be concentrated along short unstable POs of the corresponding chaotic classical system, and the quantum state bears an imprint of the PO -- a ``quantum scar''~\cite{Heller, Kaplan}. The scarring of a single-particle wave function is one of the most striking phenomena in the field of quantum chaos~\cite{quantum_chaos}. The notation of quantum scarring was introduced by one of the present authors in Ref~\cite{Heller}. Nowadays, quantum scars have been reported in a diverse range of experiments~\cite{scars_in_microwave_cavities, scars_in_optical_cavities, scars_in_quantum_wells} and simulations~\cite{scars_graphene_flakes, scars_ultra_cold_gases, relativistic_quantum_scars}. Furthermore, an effect called ``quantum many-body scarring'' has been hypothesized~\cite{MB_scars_theory_1, MB_scars_theory_2} to cause the unexpectedly slow thermalization of cold atoms, observed experimentally~\cite{MB_scars_experiment}. 

In this Letter, we describe a new kind of quantum scarring present in a 2D anisotropic HO disturbed by local perturbations such as impurity atoms. In this case, the scars are formed around the Lissajous orbits of the corresponding unperturbed system. In particular, we demonstrate that the geometry of the observed scars depend, in a similar manner as classical POs, on the frequency ratio of the confinement potential, but unlike the POs in the classical system, the scars show resilience against the alteration of the confinement. We explain our findings by generalizing the mechanism of recently discovered perturbation-induced (PI) quantum scarring~\cite{Luukko,Luukko_scars, controllability}. We also consider schemes for observing these quantum scars experimentally.


In the following, all values and equations are given in atomic units (a.u.). The Hamiltonian for a perturbed 2D quantum elliptical HO is determined by
\begin{equation}\label{quantum_Hamiltonian}
  H = \frac{1}{2}\big( -i\nabla + \mathbf{A} \big)^2 + \frac{1}{2} \left( \omega_x^2x^2  + \omega_y^2y^2 \right) + V_{\text{imp}}.
\end{equation}
The magnetic field~$\mathbf{B}$ is assumed to be oriented perpendicular to the 2D plane and incorporated via the vector potential $\mathbf{A}$.
The characteristic frequencies of the harmonic confinement are described as $\omega_x = p\omega_0$ and $\omega_y = q \omega_0$, and, for convenience, we set $\omega_0$ to unity. The perturbation $V_{\textrm{imp}}$ is modeled as a sum of Gaussian bumps with amplitude $M$ and width $\sigma$; that is, 
\begin{displaymath}
V_{\text{imp}}(\mathbf{r}) = M\sum_{i} \exp\Bigg[ -\frac{\vert \mathbf{r} - \mathbf{r}_i\vert^2}{2\sigma^2}\Bigg].
\end{displaymath}
We consider the case where the bumps are scattered randomly with a uniform mean density of two bumps per unit square. In the energy range considered here, $E = 50,\dots,250$, hundreds of bumps exist in the classically allowed region. The full width at half maximum of the Gaussian bumps $2\sqrt{2\ln2}\sigma$ is 0.235, comparable to the local wavelength of the eigenstates considered. The amplitude of the bumps is set to $M=4$, which causes strong scarring in the studied energy regime.

The Schr{\"o}dinger equation for the Hamiltonian in Eq.~(\ref{quantum_Hamiltonian}) is solved by utilizing the \texttt{itp2d} code~\cite{itp2d} based on the imaginary time propagation method. However, before considering the quantum solutions of the perturbed HO, we briefly discuss the unperturbed system, both classical and quantum.

First, we consider classical POs in an anisotropic HO without a magnetic field.
In the following, the notation $(p, q)$ refers to the frequency ratio $\omega_x/\omega_y = p/q$. Closed curves exist only if the frequencies are commensurable; i.e., the ratio $\omega_x/\omega_y$ is rational. In our notation, this occurs when $p$ and $q$ are relative primes, and the corresponding closed curves are Lissajous orbits. Geometrically, the particle has returned exactly to its starting position with its original velocity after making $p$ and $q$ oscillations between the $x$ and $y$ turning points, respectively. On the other hand, if the frequencies are incommensurable, the motion is quasiperiodic, resulting in ergodic behavior on a torus~\cite{Strogatz}. 

On the quantum side, the unperturbed system is likewise analytically solvable.
The eigenstates of an anisotropic HO can be expressed~\cite{quantum_solution} as
\begin{equation}\label{quantum_solution}
\Psi_{n,m} (x,y) = \mathcal{N} H_{n}(\sqrt{\omega_x}x) H_{m}(\sqrt{\omega_y}y)
e^{-\frac{1}{2}( \omega_x x^2 + \omega_y y^2)},
\end{equation}
where $\mathcal{N}$ is a normalization constant and $H_m(\cdot)$ is the Hermite polynomial of order $m$. The corresponding energy spectrum shows degeneracies at commensurable frequencies.


In general, the solutions of an anisotropic HO can be also examined analytically under a perpendicular magnetic field~\cite{charged_AHO}, although here we focus on the zero-field case. In addition, we want to emphasize the fact that the quantum solutions presented in Eq.~(\ref{quantum_solution}) have rectangular symmetry, even in the limit of large quantum numbers. Hence, the eigenstates in Eq.~(\ref{quantum_solution}) do not show any features of classical POs. In order to describe a classical particle, one can construct~\cite{CS_Schrodinger} a coherent state for a one-dimensional HO, more precisely, a wave packet whose center follows the corresponding classical motion. Generalized to 2D, the Schr\"{o}dinger coherent state must be a wave packet with its center mimicking a classical trajectory. This idea has been employed to form stationary coherent states reflecting the classical Lissajous orbits in terms of the time-dependent Schr\"{o}dinger coherent states~\cite{stationary_CS}. Furthermore, coherent states of this kind have been theoretically exploited to reconstruct the experimental laser modes localized on Lissajous orbits as a superposition of the HG modes~\cite{CS_and_Lissajous_curves}. Nevertheless, this artificial reconstruction of laser modes cannot explicitly manifest the quantum-classical correspondence stemming from the Schr{\"o}dinger equation.

\begin{figure}
  \centering
  \includegraphics[width=\linewidth]{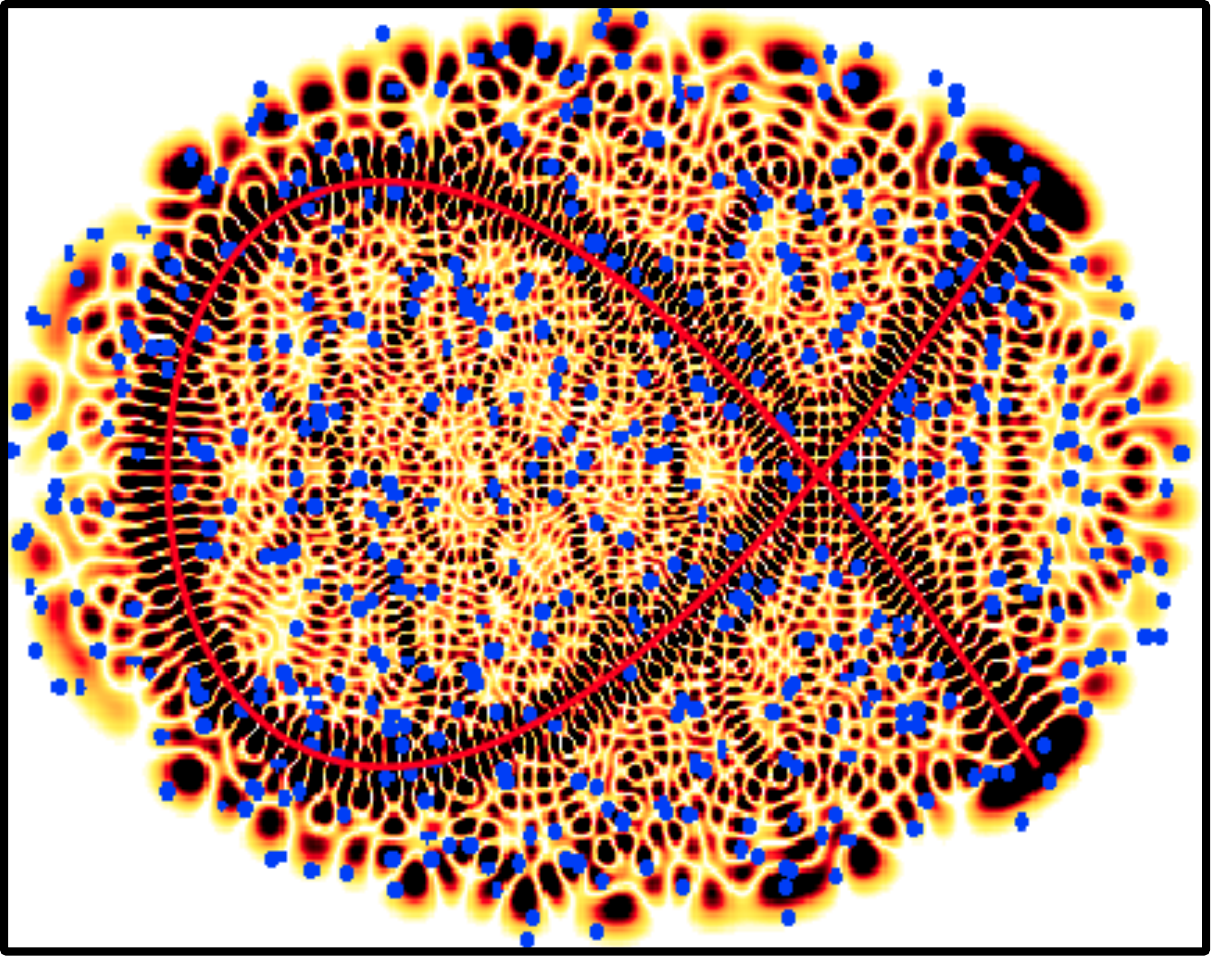}
  \caption{Alpha scar visible in the probability density of the eigenstate $n= 3453$ in an elliptical harmonic potential $(1, 2)$ perturbed by Gaussian-like bumps. The state is strongly scarred by the alpha-shape Lissajous orbit of the corresponding unperturbed potential represented as a solid red line. Blue markers denote the locations of the bumps. It is noteworthy that multiple bumps are located on the scar path.}\label{alpha_scar}
\end{figure}

When perturbed by randomly positioned Gaussian-like bumps, some of the high-energy eigenstates of the anisotropic HO are strongly scarred by Lissajous orbits of the unperturbed system. Figure~\ref{alpha_scar} shows an example of a strong quantum scar resembling the corresponding alpha-shape Lissajous orbit in the classical, unperturbed potential with commensurable frequencies $(2, 3)$. Furthermore, the presented alpha scar is counterintuitively oriented so that it maximizes the overlap with the bumps (see below for details).

\begin{figure}
  \centering
  \includegraphics[width=\linewidth]{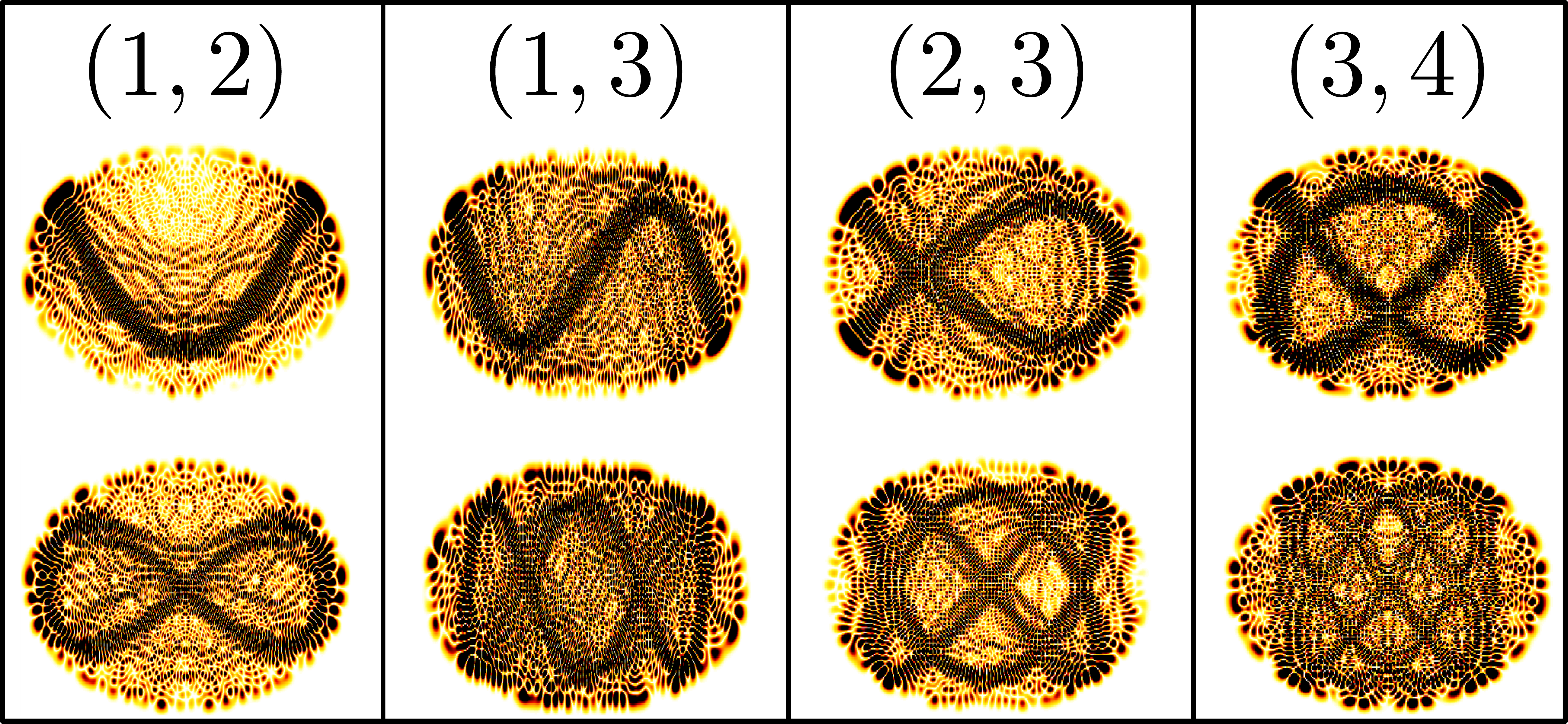}
  \caption{Examples of Lissajous scars in a two-dimensional anisotropic 
harmonic oscillator with commensurable frequencies perturbed by potential bumps. The geometries of the scars depend on the confinement potential $(p,q)$, which also defines the shape of the POs in the unperturbed system. At a fixed $(p,q)$, the scars can be divided into two subgroups: strings (upper row) and loops (lower row).}\label{scar_collection}
\end{figure}

Generally, strong quantum Lissajous scars are observed at commensurable frequencies $(p, q)$, where short classical POs exist. Examples of these quantum Lissajous scars are presented in Fig.~\ref{scar_collection}. In addition to the example cases shown in Fig.~\ref{scar_collection}, we also observe Lissajous scars related to higher commensurable frequencies $(p, q)$ such as $(2,5)$, $(3,5)$, or $(4,5)$. The eigenstate number varies between $500$ and $3900$. At given commensurable frequencies $(p,q)$, the scars appear in two distinct shapes due to the anisotropy of the oscillator: the enhanced probability distribution related to a scar either resembles an open string or a continuous loop, thus are called strings and loops, respectively. 

We stress that the Lissajous scars are not a rare occurrence at commensurable frequencies~\cite{scar_detector}; the proportion of strongly scarred states among all the first 4000 eigenstates varies from $10 \, \%$ to $60 \, \%$ at amplitude $M = 4$. Furthermore, some eigenstates contain a trace of two scars, e.g., a combination of two strings or a string and a loop. In addition to Lissajous scars, we observe quantum states that show features of classical ``bouncing-ball-like'' motion.  

As the bump density is decreased, the eigenstates of the perturbed system begin to gain traces of rectangular symmetry stemming from the unperturbed system. On the other hand, if the bump density is increased, the scars fade into completely delocalized states. The same effect is observed in the variation of the bump amplitude and width.
However, the Lissajous scars show persistence toward a modulation of the confinement, i.e., a deviation from the commensurable frequencies, as shown and analyzed below.

\begin{figure}
  \centering
  \includegraphics[width=\linewidth]{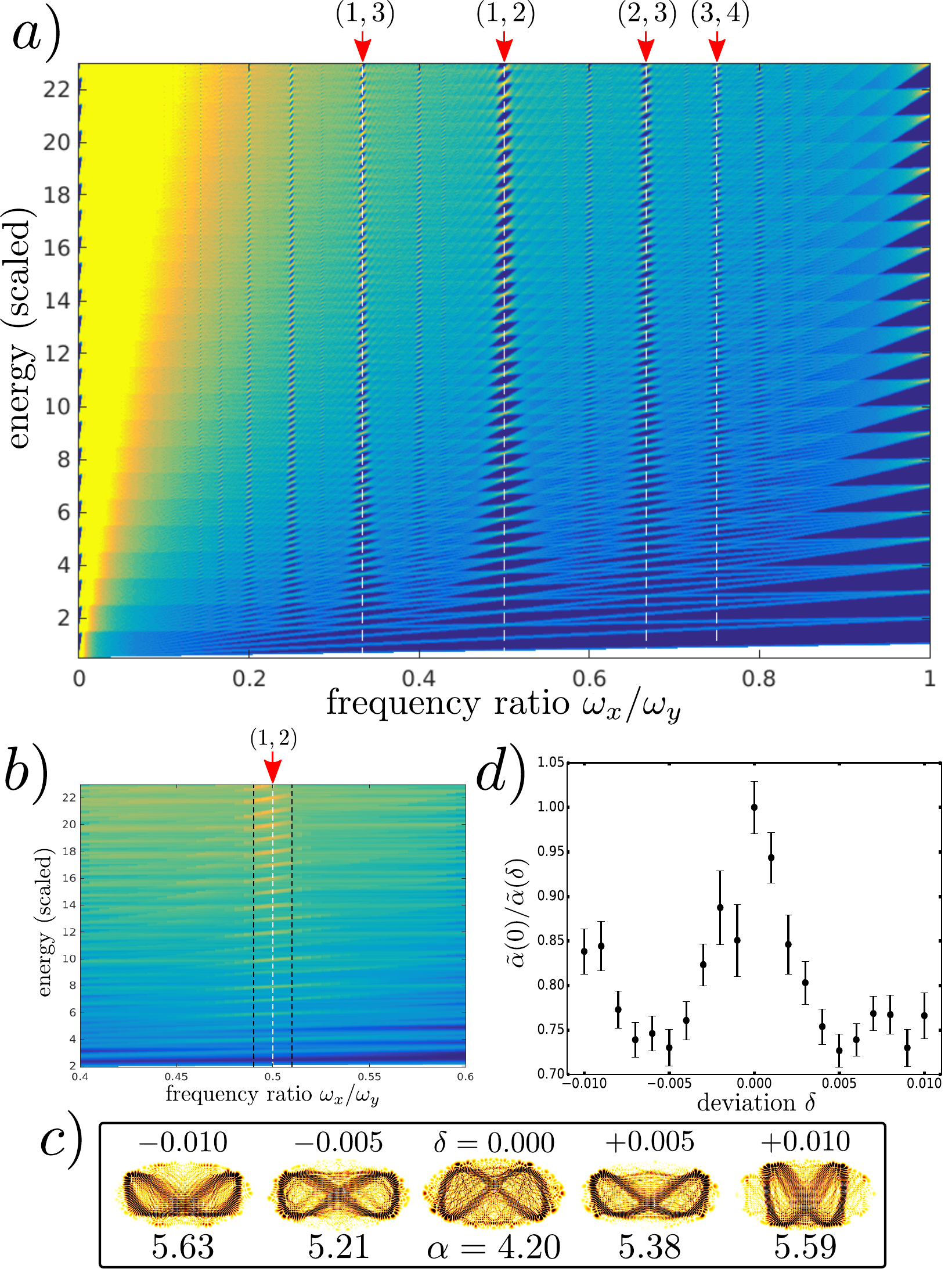}
  \caption{{\bf (a)} Density of states of the \emph{unperturbed} two-dimensional harmonic oscillator as a function of anisotropy parameter $\omega_x/\omega_y$. The dashed vertical lines indicate the commensurable $(p,q)$ that correspond to a significant abundance of scarred eigenstates in the perturbed case (see Fig.~\ref{scar_collection}). Two distinct limits are also seen in (a), namely, the unbounded case ($\omega_x/\omega_y \rightarrow 0$) and the isotropic oscillator ($\omega_x/\omega_y = 1$). {\bf (b)} Density of states of the corresponding \emph{perturbed} system as a function frequency ratio in the neighborhood of the commensurable frequency $(1, 2)$ demonstrating that the bumps are sufficiently weak enough not to fully destroy the (near) degeneracy of the unperturbed system. {\bf (c)} Examples of Lissajous scars in the vicinity of the commensurable frequency $(1,2)$ labeled with the value $\delta$ describing the deviation from the ideal frequency ratio $\omega_x/\omega_y = 0.5$. The scarring level of the quantum state is estimated by the $\alpha$ value. Note that the scars exist, although the corresponding unperturbed classical PO does not. {\bf (d)} Normalized average of $\alpha$ value as a function of the deviation $\delta$. The scarring weakens as the deviation $\delta$ increases according to the normalized average, as well as the $\alpha$ value of the individual example scars in (c).
  }\label{Lissajous_DOS}
\end{figure}

To further analyze the Lissajous scarring, we compute the density of states (DOS) as a sum of the states with a Gaussian energy window of $0.001$ a.u. Figure~\ref{Lissajous_DOS} visualizes the DOS for a few thousand lowest energy levels as a function of the ratio $\omega_x/\omega_y$. Figure~\ref{Lissajous_DOS}(a) corresponds to an unperturbed system, and the dashed vertical lines mark the accidental degeneracies at the rations $(p,q)$ shown in Fig~\ref{scar_collection}. The proportion and strength of scarred states depend on the degree of degeneracy in the unperturbed spectrum: more and stronger scars appear when more energy levels are (nearly) crossing. Figure~\ref{Lissajous_DOS}(b), on the other hand, illustrates for the commensurable frequency $(1,2)$ that the perturbation caused by the bumps is sufficiently weak enough not to completely destroy this degeneracy structure.

We supplemented the scar analysis by introducing a localization measure ($\alpha$ value) for a normalized eigenstate $n$ defined as
$
\alpha_n  = Z\int \vert \psi_n (\textbf{r}) \vert^4 \, \textrm{d} \textbf{r},
$
where the normalization factor $Z$ is determined by the classical area for the energy $E_n$ in the unperturbed system~\cite{IPR}. As the $\alpha$ value describes the localization of the probability density of a state, we employ it here to estimate qualitatively the strength of scarring. 

If the confinement deviates from a commensurable frequency $(p,q)$ while keeping $V_{\textrm{imp}}$ otherwise unchanged, the scars persist. Figure~\ref{Lissajous_DOS}(c) presents examples of strong, looplike Lissajous scars in the neighborhood of the commensurable frequency $(1,2)$, marked with the deviation $\delta$ from the corresponding frequency ratio $\omega_x/\omega_y = 0.5$. We want to emphasize that the classical POs that the scars resemble do not exist in the perturbed or even in the unperturbed system when the frequency ratio $\omega_x/\omega_y$ differs from the commensurable frequency $(1,2)$. Although scarred states exist outside the optimal frequency ratio, the strength of the scarring decreases as indicated by the $\alpha$ value of the scars shown in Fig.~\ref{Lissajous_DOS}(c). 

For a more complete picture, we also compute an average $\tilde{\alpha}(\delta)$.
More precisely, we consider 30 looplike Lissajous scars, such scars as in Fig.~\ref{Lissajous_DOS}(c), at different deviations in the interval $\vert \delta \vert = 0.01$ indicated by the black vertical lines in Fig.~\ref{Lissajous_DOS}(b). The normalized average $\tilde{\alpha}(0)/\tilde{\alpha}(\delta)$ shown in Fig.~\ref{Lissajous_DOS}(b) reveals that the scarring becomes weaker as the deviation $\delta$ from the commensurable frequency increases. Along with the average scarring strength, the number of scars reduces with increasing deviation. In practice, the scars connected to the commensurable frequency $(1,2)$ have vanished outside the deviation interval presented in Fig.~\ref{Lissajous_DOS}(b). However, both effects can be compensated at a certain level by adjusting the perturbation.   

Before delving into the mechanism behind these oddly ordered structures, we want to address two aspects. First, the considered amplitude of the bumps ($M = 4$) is small in comparison to the total energy, making each individual bump a small perturbation. Nonetheless, together the bumps form sufficient perturbation enough to destroy classical long-time stability; any stable structures present in the otherwise chaotic Poincar\'{e} surface of section are minuscule compared to $\hbar = 1$.

Second, the Lissajous scars cannot be explained by dynamical localization~\cite{dynamical_localization_I,dynamical_localization_II}: it corresponds to localization in angular momentum space, whereas the scars are localized in position space. In addition, dynamical localization is not able to explain that scars generally orient to coincide with as many bumps as possible (see also Refs.~\citep{Luukko, controllability}). Furthermore, even though similar in appearance, the conventional scar theory~\cite{Heller, Kaplan,Linear_scar_theory, Linear_and_nonlinear_scar_theory} cannot describe the Lissajous scarring, as it would require the existence of short, moderately unstable POs in the perturbed system.

To explain the Lissajous scarring, we generalize the PI scar theory beyond circularly symmetric potentials~\cite{Luukko, Luukko_scars,controllability}. Recently, PI scars have drawn attention since they have been demonstrated to be highly controllable~\cite{controllability}, and can be utilized to propagate quantum wave packets in the system with high fidelity~\cite{Luukko}. Combined, this may open a door to coherently modulate quantum transport in nanoscale devices by exploiting the scarring. In addition, the PI scars have been analyzed~\cite{PI_scars_and_statistics} in the framework of quantum chaos. Furthermore, the PI scarring is expected to be manifested in a dense random gas as a polyatomic trilobite Rydberg molecule~\cite{Trilobite}.
 
For PI scars to occur, we only require two ingredients: the existence of special (nearly) degenerate states called a ``resonant'' set in the unperturbed system, and the individual bumps need to have a short spatial range. Hence, we extend the PI scarring mechanism to hold for a larger set of systems with a lower symmetry than circular symmetry~\citep{Luukko, Luukko_scars, controllability, PI_scars_and_statistics}, such as an anisotropic oscillator. 

In an anisotropic oscillator, the resonant sets stem from the accidental degeneracy occurring at commensurable frequencies; e.g., the dashed lines in Fig.~\ref{Lissajous_DOS}(a) correspond to frequency ratios with substantial degeneracy. These resonant sets are related to a family of classical POs, which ensures that some linear combinations of the states in a resonant set are scarred by Lissajous orbits.

A \emph{moderate} perturbation forms eigenstates that are linear combinations of a single resonant set. Based on the variational theorem, the states corresponding to extremal eigenvalues extremize the perturbed Hamiltonian. Because the states in a resonant set are (nearly) degenerate, this basically means extremizing the perturbation. In the extremization, the system prefers the scarred states since the bumps causing the perturbation are \emph{localized}~\cite{character_of_bumbs}. Thus, scarred states can effectively maximize (minimize) the perturbation by selecting paths coinciding with as many (few) bumps as possible. As a result, the extremal eigenstates arising from each resonant set often contain scars of the corresponding PO.

The elliptical oscillator has also experimental relevance: it realistically models disordered quantum with soft boundaries. Thus, it provides a platform, as a quantum counterpart of classical billiard, to investigate the nature of quantum chaos, e.g., with a statistical analysis of the energy levels~\cite{quantum_chaos}. 

An important avenue of future research is to analyze the effect of PI scarring on the conductance of the QD in more detail (see Refs.~\cite{Luukko, controllability}) by employing realistic quantum transport calculations. Previous studies (see, e.g., Refs.~\cite{scars_graphene_flakes,scars_and_conductance}) have shown that the effect of (conventional) scarring can be observed in the conductance fluctuations. Moreover, open QDs are suitable for wave function imaging based on shifts in the energy of the single-particle resonances, induced by an AFM tip~\cite{wavefunction_imaging_I,wavefunction_imaging_II,wavefunction_imaging_III}. In addition, the scarred eigenstates of an electron in a QD may be measured with quantum tomography~\cite{quantum_tomography}. For completeness, we want to address that a PI scar can be even created by a single bump, generated in a controlled manner by, e.g., a conducting nanotip~\cite{nanotip}.

Outside of QDs, we suggest that Lissajous scars may be possible to detect in optical systems, frequently employed to observe conventional quantum scars (see, e.g., Refs.~\cite{scars_optical_systems_I,scars_optical_systems_II,scars_optical_systems_III}),and to study quantum chaos in general~\cite{quantum_chaos}. For some types of polarization, the three components of the electric field decouple, and thereby, for example, a quasimonochromatic light can be described in terms of a scalar wave equation~\cite{LLP}. Further, in the paraxial approximation (at the lowest order), the slowly variating amplitude of the field formally satisfies a single-particle Schr{\"o}dinger equation in a dielectric medium with spatially dependent refractive index~\cite{paraxial_approximation_I, paraxial_approximation_II,optical_schrodinger_equation}. Thus, the formulation allows us to interpret the light propagation as the evolution of a massive particle~\cite{optical_schrodinger_equation,light_propagation_I,light_propagation_II, light_propagation_III}, and Schr{\"o}dinger-like behavior, such as scarring, should emerge. In particular, with a suitable choice of the refractive index, this ``optical Schr{\"o}dinger equation'' (see, e.g., Ref.~\cite{optical_schrodinger_equation}) reduces to an anisotropic HO, such as arising from the quantum Hamiltonian~(\ref{quantum_Hamiltonian}) without a magnetic field. The potential bumps may be realized by creating small, localized deviations of the refractive index, which can be even randomly positioned. Therefore, optical fibers~\cite{paraxial_approximation_II,optical_fibers} may be employed to experimentally investigate PI scars, along with other quantum phenomena.

In conclusion, we have shown that a two-dimensional anisotropic harmonic oscillator supports quantum scars induced by randomly scattered potential bumps. These quantum Lissajous scars are relatively strong, and their abundance and geometry are related to commensurable frequencies. This counterintuitive phenomenon emerges from the extended concept of PI scarring as a combination of resonant sets and the localized nature of the perturbation. We also considered the experimental consequence of the quantum Lissajous scars. In particular, an optical approach may indicate a path to experimentally realize these scars in optical fibers by utilizing the analogy between the quantum theory and classical electromagnetism. Lissajous scars are hence a peculiar example of quantum suppression of classical chaos not only for establishing a relationship between quantum states and classical POs in the 2D anisotropic harmonic oscillator, but also for optics.

\begin{acknowledgments}
We are grateful to Janne Solanp{\"a}{\"a}, Matti Molkkari, and Rostislav Duda for useful discussions. We also acknowledge CSC -- Finnish IT Center for Science for computational resources. Furthermore, J. Keski-Rahkonen thanks the Magnus Ehrnrooth Foundation for financial support.
\end{acknowledgments}


\begin{thebibliography}{99}
\bibitem{scar_detector} We have developed a tool for the automated detecting and classification of PI scars employing machine learning along with different approaches for measuring the strength of a scar. However, this will be considered in detail in another publication.
\bibitem{character_of_bumbs} Note that the statement does not assert if the bumps are repulsive or attractive. Although we have focused on the repulsive bumps, PI scars persist also in perturbed potential landscapes, including attractive bumps. The PI scars caused by attractive bumps are confirmed by our simulations, and they are well understood within the generalized scarring mechanism. Furthermore, the statement does not quantify the localization of the bumps for PI scarring to occur; the details depend on the given system, e.g., on the density, width, and amplitude of the bumps along with the level of the near degeneracy in the unperturbed quantum system.
\end{thebibliography}
\end{document}